\documentclass[a4paper,11pt]{article}
\pdfoutput=1 

\usepackage{jheppub} 

\usepackage[T1]{fontenc} 

\usepackage{slashed}
\usepackage{bm}
\usepackage{upgreek}
\usepackage{mathrsfs}
\usepackage{times}
\usepackage{amsthm}
\usepackage{amssymb}
\usepackage{epsfig}
\usepackage{graphicx}
\usepackage{amsmath}

\newcommand{\barray}{\begin{eqnarray}}
\newcommand{\earray}{\end{eqnarray}}

\newcommand{\beq}{\begin{equation}}
\newcommand{\eeq}{\end{equation}}
\newcommand{\ba}{\begin{array}}
\newcommand{\ea}{\end{array}}
\newcommand{\bea}{\begin{eqnarray}}
\newcommand{\eea}{\end{eqnarray} }
\newcommand{\be}{\begin{eqnarray}}
\newcommand{\ee}{\end{eqnarray} }
\newcommand{\bal}{\begin{align}}
\newcommand{\eal}{\end{align}}
\newcommand{\bi}{\begin{itemize}}
\newcommand{\ei}{\end{itemize}}
\newcommand{\ben}{\begin{enumerate}}
\newcommand{\een}{\end{enumerate}}
\newcommand{\bc}{\begin{center}}
\newcommand{\ec}{\end{center}}
\newcommand{\bt}{\begin{table}}
\newcommand{\et}{\end{table}}
\newcommand{\btb}{\begin{tabular}}
\newcommand{\etb}{\end{tabular}}

\DeclareOldFontCommand{\brianup}{\upshape}{\mathrm}

\DeclareSymbolFont{EUR}{U}{eur}{m}{n}
\SetSymbolFont{EUR}{bold}{U}{eur}{b}{n}
\DeclareSymbolFontAlphabet{\eur}{EUR}

\DeclareSymbolFont{EUB}{U}{eur}{b}{n}
\SetSymbolFont{EUB}{bold}{U}{eur}{b}{n}
\DeclareSymbolFontAlphabet{\eub}{EUB}

\DeclareSymbolFont{AMSb}{U}{msb}{m}{n}
\DeclareSymbolFontAlphabet{\mathbb}{AMSb}

\newcommand{\notyet}[1]{{}}

\DeclareMathSymbol{\varGamma}{\mathord}{letters}{"00}
\DeclareMathSymbol{\varDelta}{\mathord}{letters}{"01}
\DeclareMathSymbol{\varTheta}{\mathord}{letters}{"02}
\DeclareMathSymbol{\varLambda}{\mathord}{letters}{"03}
\DeclareMathSymbol{\varXi}{\mathord}{letters}{"04}
\DeclareMathSymbol{\varPi}{\mathord}{letters}{"05}
\DeclareMathSymbol{\varSigma}{\mathord}{letters}{"06}
\DeclareMathSymbol{\varUpsilon}{\mathord}{letters}{"07}
\DeclareMathSymbol{\varPhi}{\mathord}{letters}{"08}
\DeclareMathSymbol{\varPsi}{\mathord}{letters}{"09}
\DeclareMathSymbol{\varOmega}{\mathord}{letters}{"0A}

\theoremstyle{plain}

\theoremstyle{definition}

\theoremstyle{remark}

\makeatletter\@addtoreset{equation}{section}
\makeatother

\title{{\boldmath Gauge theory of Lorentz group as a source of the dynamical electroweak symmetry breaking}\\
{\small \it dedicated to the memory of D.I.Diakonov}}

\author{M.A.Zubkov}

\affiliation{Institute for Theoretical and Experimental Physics\\B.Cherjemushkinskaya - 25, Moscow, Russia}

\emailAdd{zubkov@itep.ru}

\abstract{
We consider the gauge theory of Lorentz group coupled in a nonminimal way to fermions.
We suggest the hypothesis that the given theory may
exist in the phase with broken chiral symmetry and without confinement. The lattice discretization of the model is described. This unusual strongly coupled theory may appear to be the source of the dynamical electroweak symmetry breaking. Namely, in this theory all existing fermions interact with the $SO(3,1)$ gauge field. In the absence of the other interactions the chiral condensate may appear and all fermionic excitations may acquire equal masses. Small corrections to the gap equations due to the other interactions may cause the appearance of the observed hierarchy of masses.

}

\begin{document}
\maketitle
\flushbottom

\section{Introduction}

The scalar excitation recently found at $125$ GeV is now interpreted as the Higgs boson \cite{CMSHiggs, ATLASHiggs}. The existence of the other Higgs bosons with the same (or, larger) production cross - sections is excluded within the wide ranges of masses approximately from $130$ GeV to $550$ GeV \cite{CMSexc,ATLASexc}. However, this does not mean that the existence of scalar particles within these ranges of masses is excluded completely. The particles with smaller values of production cross - sections are allowed. In \cite{VZ2013,VZ2012,VZ2013_2} it was suggested that such particles may appear. It was supposed that these particles together with the $125$ GeV Higgs may be composed of known Standard Model fermions due to the unknown strong interaction between them (with the scale $\Lambda$ above $1$ TeV). This unknown interaction if it exists should have very specific properties that make it different from the conventional technicolor (TC) interactions \cite{Technicolor,Technicolor_,Technicolor__} (for a recent alternative to the technicolor see, for example, \cite{Simonov}).

First, these interactions cannot be confining since otherwise they would confine quarks and leptons to the extremely small regions of space $\sim 1/\Lambda$, so that all strong and weak interaction physics would be missed. At the same time these interactions should provide the spontaneous chiral symmetry breaking needed to make W and Z bosons massive. Some models with such properties  were already discussed in the framework of the topcolor \cite{topcolor1,topcolor,Marciano:1989xd,cveticreview,Miransky,modern_topcolor,Simmons,Hill:1996te,Buchalla:1995dp}. For the review of the conventional technicolor see  \cite{Weinberg:1979bn,Susskind:1978ms,Simmons,Chivukula:1990bc}. The topcolor assisted technicolor (combines both technicolor and topcolor ingredients) was considered in \cite{Hill:1994hp, LaneEichten, Popovic:1998vb, Braam:2007pm,Chivukula:2011ag}.
The models based on the extended color sector were considered in \cite{top_coloron,top_coloron1}. For the top - seesaw see  \cite{topseesaw}.
 In \cite{Simonov,VZ2013,VZ2012} the scenario was suggested according to which the chiral condensates appear corresponding to some of the Standard Model fermion fields (not necessarily of the top quark only). This pattern provides both masses for the W, Z bosons and for the fermionic particles.

  The considerations of \cite{VZ2013,VZ2012,VZ2013_2} were based on the NJL approximation. Due to the non - renormalizability it is  to be considered as the phenomenological model with the finite ultraviolet cutoff $\Lambda$. In the majority of papers on the NJL approximation only the leading order in $1/N$ expansion was considered. The contributions of higher loops, however, may be strong.  Actually, nonperturbative lattice results and higher orders of the perturbation theory demonstrate that all dimensional parameters of the relativistic NJL models  are of the order of the cutoff \cite{cveticreview,cvetic,latticeNJL}. Moreover, the higher loop corrections may destroy the dynamical electroweak symmetry breaking (DEWSB) just like the quadratic divergent contributions to the Higgs  boson mass destroy the DEWSB in the Standard Model. That's why  the problem of higher order corrections in the NJL models is related to the hierarchy problem of the Standard Model.

In \cite{VZ2012} it was proposed to look at the effective four - fermion (NJL) approximation as at the effective low energy theory, in which only the one - loop results have sense. The higher loops are simply disregarded.
This approach is similar to the consideration of the hydrodynamics with the corrections due to phonon loops disregarded.
In quantum hydrodynamics there exist the divergent contributions to physical quantities  (say, to the vacuum energy). If these divergences are omitted, the classical hydrodynamics appears as a low energy approximation to quantum theory. Why is this so? The quantum hydrodynamics is an effective low energy theory with the cutoff $\Lambda$. The complete theory contains the trans - $\Lambda$ degrees of freedom, whose contribution to vacuum energy exactly cancels the divergent loop phonon contributions of quantum hydrodynamics. This occurs  due to the thermodynamical stability of vacuum \cite{quantum_hydro}. It was also suggested that the similar mechanism may be responsible for the cancellation of the ultraviolet divergences in quantum gravity as well as for the cancellation of the dangerous quadratically divergent contribution to the Higgs boson mass in the Standard Model \cite{hydro_gravity}.

It is worth mentioning that in QCD \cite{NJLQCD} and in Technicolor \cite{Simmons} usually only the one - loop NJL results are taken into account. In many papers on the NJL models of top quark condensation (TC) \cite{Miransky,topcolor1} the cutoff was assumed to be many orders of magnitude larger than the electroweak scale, but the consideration was limited to one loop. Next to leading approximation in TC was considered in \cite{cvetic,cveticreview}. In the Extended Technicolor (ETC) \cite{Simmons} the effect of the ETC gauge group is taken into account through the effective four - fermion term. Loop contributions would give values of masses $\sim \Lambda^2_{ETC}, \Lambda_{ETC} \gg M_T$, but these contributions are usually omitted.

In \cite{VZ2012} it was proposed that in the models of the dynamical electroweak symmetry breaking with composite Higgs bosons there exist the contributions of the microscopic theory due to the trans - $\Lambda$ degrees of freedom that cancel the dominant higher loop divergences of the low energy NJL effective theory. Therefore, the one - loop results dominate. In the present paper we follow this proposition. Sure, this is a kind of the fine - tuning that is often considered as unnatural. But, is this possible to avoid the fine - tuning while providing the fermion masses from less than $1$ eV for neutrino to about $174$ GeV for the top - quark (the difference is eleven orders of magnitude)?

It was proposed \cite{VZ2012,VZ2013,VZ2013_2} that in the zero order approximation all fermions have the same masses. Small corrections to the gap equations are able to provide considerable difference between the quark masses. This is how the hierarchy of masses appears. Unfortunately, in addition, in this approach numerous light higgs bosons appear. Some physics is to be added to \cite{VZ2012,VZ2013,VZ2013_2} in order to provide extra light higgs bosons with the masses on the order of the top quark mass (see discussion below in section 5).  The given pattern prompts that the mentioned unknown forces binding quarks are to be universal, i.e. they have to act in the same way on all fermions. We know only one kind of forces that act in the same way on all particles: the gravitational forces. At the same time, the scale of the Riemannian gravity is the Plank mass that is many orders of magnitude larger than the expected scale for these interactions. However, one may consider quantum gravity with torsion. Its dynamical variables are the vierbein (that is related to Riemannian gravity) and the $SO(3,1)$ connection that gives rise to torsion. With the vierbein frozen we come to the gauge theory of Lorentz group \cite{Minkowski,Minkowski_}. At the present moment we cannot point out any particular  quantum gravity theory that gives rise to the dynamical torsion with the scale much smaller than the scale of the fluctuating metric. (The former scale is not fixed but is assumed to be between $10^3$ TeV and $10^{15}$ GeV while the  latter scale is assumed to be given by the Plank mass $\approx 10^{19}$ GeV.)

In this paper we suggest the possibility that the gauge theory of Lorentz group binds quark and lepton fields giving rise to the dynamical electroweak symmetry breaking in the spirit of the models of top - quark condensation \cite{topcolor1, topcolor} and the models considered in \cite{Simonov}. This theory is strongly coupled but it has the properties that make  it different from the QCD - like technicolor theories \cite{Technicolor,Technicolor_,Technicolor__}. Therefore, we expect that it may provide chiral symmetry breaking without the confinement.
One of the characteristic features of this theory is that the term linear in curvature appears similar to the Palatini action of the Poincare gravity. When the other terms for the gauge field are neglected, this term gives rise to the four - fermion interactions.  In the context of Poincare gravity (i.e. in the presence of nontrivial vierbein) this was considered in
\cite{Freidel:2005sn,Randono:2005up,Mercuri:2006um,Alexandrov}. The induced four -
fermion interactions were considered in
\cite{Xue,Alexander1,Alexander2} as a source of the fermion condensation used,
mainly, in a cosmological background. In our previous paper \cite{Z2010_3} we
have suggested that nonminimal coupling \cite{Shapiro} of fermion fields to torsion (when Parity is broken by the torsion action)
 may provide the condensation of the fermion bilinears composed of the right - handed fermions,
and cause the Dynamical
Electroweak Symmetry Breaking (DEWSB) if the technifermions are placed in the right - handed spinors while the Standard Model fermions are placed in the left - handed spinors.
In the present paper we do not consider such a pattern and do not introduce the technifermions at all.

It is worth mentioning that the most general form of the nonminimal coupling of fermions to gauge field of Lorentz group (the gravity with torsion)
was suggested only recently in \cite{Alexandrov}. Later it was used, for example, in \cite{Diakonov}. It is important that dealing with the Lorentz gauge theory we cannot consider only the minimal coupling because the nonminimal interaction between fermions and the gauge group is generalted dynamically. Therefore, following \cite{Diakonov} we consider the nonminimal coupling from the very beginning.
It is worth mentioning that the parity violation appears in the model of the fermions coupled  to gravity with torsion in two cases:
if there is the nonminimal parity breaking coupling of fermions to gravity (this was considered, for example, in \cite{Alexandrov}) and  if the pure gravity action contains the parity - breaking terms (for example, in \cite{Freidel:2005sn} the Holst action is considered that breaks parity).
In both cases there take place effects of atomic parity violation.
However, in the model considered in the present paper neither of the two mentioned above sources of
parity violation appear:
the considered non-minimal coupling of fermions to gravity does not break parity, while the Holst term of the gravity action is absent; also
there are no parity breaking terms of the action quadratic in curvature.

Strongly coupled theories are to be investigated using nonperturbative methods. The most powerful nonperturbative method is the numerical simulation of the model in lattice discretization. Therefore, we consider the lattice discretization. In its essence our discretization is similar to that of suggested in \cite{Z2006} for the Poincare gravity, and in \cite{Diakonov} for the  theory of quantum gravity suggested by D.I.Diakonov.

The paper is organized as follows. In the 2-nd section we consider fermion
fields coupled to the gauge theory of the Lorentz group.  In the 3-rd section we consider the action
for the gauge field and derive the four - fermion interactions that appear after the integration over
torsion. In the $4$ - th section we apply NJL technique to the four -
fermion interactions of fermions and describe how the chiral symmetry
breaking occurs. In the $5$ - the section we consider the model that appears when small perturbations to the exchange by the Lorentz group gauge bosons are taken into account. In section $6$ we introduce the lattice discretization needed for the numerical investigation of the model.   In section $7$ we end with the conclusions.

\section{Fermions coupled to the gauge field of Lorentz group}

We consider the CP - invariant action of a massless Dirac spinor coupled to the $SO(3,1)$ gauge field in
the form \cite{Alexandrov}:

\begin{eqnarray}
S_f & = & \frac{i}{2}\int  \{ \bar{\psi} \gamma^{\mu} \zeta
D_{\mu} \psi - [D_{\mu}\bar{\psi}]\bar{\zeta}
\gamma^{\mu}\psi \} d^4 x \label{Sf}
\end{eqnarray}

Here $\zeta=\eta + i\theta$ is the coupling constant,
 the covariant derivative is $D_{\mu} =
\partial_{\mu} + \frac{1}{4}C_{..\mu}^{ab} \gamma_{[a}\gamma_{b]}$;
$\gamma_{[a}\gamma_{b]} =
\frac{1}{2}(\gamma_{a}\gamma_{b}-\gamma_{b}\gamma_{a})$. The spin
connection is denoted by $C_{\mu}$. It is related to Affine connection $\Gamma^{i}_{jk}$
and torsion $T^a_{.\mu \nu}$ as follows:
\begin{eqnarray}
 \Gamma^{a}_{\mu
\nu} &=&  C^{a}_{ . \mu\nu}\nonumber\\
T^a_{.\mu \nu} & = & C^{a}_{ .
[\mu\nu]}
\end{eqnarray}
Indices are lowered and lifted via metric $g$ of Minkowsky space as
usual.

The symmetry group of the theory is the group of local Lorentz transformations \cite{Minkowski,Minkowski_}. The infinitesimal transformations are given by
\begin{eqnarray}
x^i &\rightarrow &\Lambda^i_{.j} (x) x^j = x^i + \epsilon^{i}_{j}(x) x^j, \quad \epsilon(x) \in so(3,1)\nonumber\\
\psi(x) &\rightarrow& \tilde{\Lambda}\psi(\Lambda^i_{.j} (x) x^j)=(1-\frac{1}{4} \epsilon^{ab} \gamma_{[a}\gamma_{b]})\psi(\Lambda^i_{.j} (x) x^j),\nonumber\\
\partial_j+\frac{1}{4}C_j^{ab}(x)\gamma_{[a}\gamma_{b]}& \rightarrow &\Lambda_{.j}^{k}\tilde{\Lambda} \Bigl(\partial_k+ C^{{a}{b}}_{\bar{k}}(\Lambda^i_{.j} (x) x^j)\gamma_{[a}\gamma_{b]}\Bigr)\tilde{\Lambda}^{-1}
\end{eqnarray}
Another way to look at this symmetry is to restore the field of the vierbein that corresponds to vanishing Riemannian curvature. The symmetry group of this system is $Diff \otimes SO(3,1)_{local}$. Fixing the trivial value of the vierbein $e^a_k = \delta^a_k$ we break this symmetry: $Diff \otimes SO(3,1)_{local} \rightarrow SO(3,1)_{local}$. This breakdown is accompanied by distinguishing local Lorentz transformations out of the general coordinate transformations.

Eq. (\ref{Sf}) can be rewritten as follows:
\begin{eqnarray}
S_f & = & \frac{1}{2}\int \{i\bar{\psi} \gamma^{\mu}\zeta\nabla_{\mu} \psi - i[\nabla_{\mu}\bar{\psi}]\bar{\zeta} \gamma^{\mu}\psi\nonumber\\&& + \frac{i}{4}
C_{abc}\bar{\psi}[\{ \gamma^{[a}\gamma^{b]},\gamma^c\}\eta - i[
\gamma^{[a}\gamma^{b]},\gamma^c] \theta] \psi\} d^4 x \label{Sf1}
\end{eqnarray}
Here $\nabla$ is the usual derivative. Next, we
obtain:
\begin{eqnarray}
S_f & = & \frac{1}{2}\int \{i\bar{\psi} \gamma^{\mu}{\zeta}\nabla_{\mu} \psi -
i[\nabla_{\mu}\bar{\psi}]\bar{\zeta}\gamma^{\mu}\psi\nonumber\\&& - \frac{1}{4}
C_{abc}\bar{\psi}[-2\epsilon^{abcd}\gamma^5 \gamma_d \eta +4
\eta^{c[a} \gamma^{b]} \theta ] \psi\} d^4 x \label{Sf2}
\end{eqnarray}

Now let us introduce the irreducible components of torsion:
\begin{eqnarray}
S^i& =& \epsilon^{jkli}T_{jkl}\nonumber\\
T_i& =& T^j_{.ij}\nonumber\\
T_{ijk}& =& \frac{1}{3}(T_j \eta_{ik}-T_k\eta_{ij}) -
\frac{1}{6}\epsilon_{ijkl}S^l + q_{ijk}
\end{eqnarray}

In terms of $S$ and $T$ (\ref{Sf2}) can be rewritten as:
\begin{eqnarray}
S_f & = & \frac{1}{2}\int \{i\bar{\psi} \gamma^{\mu}\eta
\nabla_{\mu} \psi - i[\nabla_{\mu}\bar{\psi}]{\eta} \gamma^{\mu}\psi\nonumber\\&& +
\frac{1}{4}\bar{\psi}[\gamma^5 \gamma_d \eta S^d  - 4\theta T^b \gamma_{b}] \psi\} d^4 x \label{Sf22}
\end{eqnarray}

One can see that $\theta$ disappears from the first term with usual derivatives. However, it remains in the second term. It is worth mentioning that in the theory with dynamical field $C^{ab}_{..\mu}$ it is not possible to consider only the minimal coupling with $\theta = 0$ because the second term with nonzero $\theta$ is to appear dynamically.

\section{Gauge field action}

Let us consider the action of the form $S_g = S_T + S_G$ with
\begin{equation}
S_T = -M_T^2 \int  G d^4x \label{ST}
\end{equation}
and
\begin{eqnarray}
S_G &=&  \beta_1 \int  G^{abcd}G_{abcd}d^4x+\beta_2 \int
G^{abcd}G_{cdab}d^4x\nonumber\\&& +\beta_3 \int G^{ab}G_{ab}d^4x+\beta_4 \int
G^{ab}G_{ba}d^4x\nonumber\\&& +\beta_5 \int G^2d^4x+\beta_6 \int
A^2 d^4x\label{ST2}
\end{eqnarray}
with coupling constants $\beta_{1,2,3,4,5,6}$. Here $G^{ab}_{..\mu\nu} = [D_{\mu},D_{\nu}]$ is the $SO(3,1)$ curvature,
$G^{abcd}=\delta^c_{\mu}\delta^d_{\nu}G^{ab}_{\mu\nu}$, $G^{ac}=G^{abc}_{...b}$, $G =
G^a_a$,
\begin{equation}
A = \epsilon^{abcd} G_{abcd} \label{A}
\end{equation}
 Actually,
Eq. (\ref{ST2}) is the most general quadratic in curvature action that does
not break Parity \cite{Diakonov2}. The topological charge may be composed of different terms of Eq. (\ref{ST2}):
\begin{equation}
{\cal Q} \sim  \int d^4x (R^2 - 4 R^{ab} R_{ab} + R^{abcd}R_{cdab})
\end{equation}
Therefore, up to the the topological term we have five independent terms quadratic in curvature.

$S_T$ may be represented in terms of torsion
 \cite{Mercuri:2006um}:

\begin{eqnarray}
S_T =  M_T^2 \int \{\frac{2}{3}T^2 - \frac{1}{24}S^2 \}d^4x + \tilde{S}
\end{eqnarray}
Here $\tilde{S}$ depends on $q$ but does not depend on $S, T$.
Let us suppose for a moment that it is possible to neglect $S_G$ in some approximation. Then the  integration over torsion degrees of freedom can be performed for the system that consists of the Dirac fermion
coupled to the gauge field. The result of this integration is \cite{Alexandrov,Freidel:2005sn}:

\begin{eqnarray}
S_{eff}& = & \frac{1}{2}\int \{i\bar{\psi} \gamma^{\mu} \eta
\nabla_{\mu} \psi - i[\nabla_{\mu}\bar{\psi}] \eta \gamma^{\mu}\psi\}d^4x\nonumber\\&& -
\frac{3}{32M_T^2} \int
\{V^2 \theta^2 - A^2\eta^2 \} d^4x \label{F42}
\end{eqnarray}
Here we have defined:
\begin{eqnarray}
V_{\mu} & = & \bar{\psi} \gamma_{\mu}  \psi \nonumber\\A_{\mu} & = &
\bar{\psi}\gamma^5 \gamma_{\mu}  \psi
\end{eqnarray}

  The next step is the consideration of the fermions coupled to the gauge field of Lorentz group
 with action $S_T+S_G$. This theory may appear to be  renormalizable due to the presence of terms quadratic
 in curvature \cite{Sesgin,Elizalde}. Our supposition
  is that the theory contains the scale  $\Lambda_{\chi} \sim M_T$. The effective charges entering the term of the action with the
 derivative of torsion depend on the ratio $\epsilon/\Lambda_{\chi}$, where $\epsilon$ is the energy scale of the considered physical process.
We suppose that under certain circumstances running coupling constants $\beta_i(\epsilon/\Lambda_{\chi})$ are decreased with the decrease of $\epsilon$. This does not contradict to the asymptotic freedom ($\beta_i(\epsilon/\Lambda_{\chi}) \rightarrow \infty $ at $\epsilon \rightarrow \infty$) that is often considered as the condition for the self - consistency of the gauge theory. However, due to the existence of $S_T$ in the action we do not expect confinement, and at small enough energies $\epsilon << \Lambda_{\chi}$ it is possible to neglect $S_G$. The theory with the low energy effective action Eq. (\ref{F42}) has the natural cutoff $\Lambda_{\chi}$. In the next section we shall demonstrate that in this theory the dynamical breakdown of chiral symmetry is possible.

\section{Chiral symmetry breaking }

As it was mentioned in the Introduction we restrict ourselves with the one - loop analysis of the effective NJL model derived in the previous secton. This is based on the assumption that due to a certain symmetry the dominant divergencies that appear at more than one loop are exactly cancelled by contribution of the trans - $\Lambda_{\chi}$ degrees of freedom of the microscopic theory. (The microscopic theory in our case is the complete gauge theory of the Lorentz group or the Poincare gravity that contains it, or some other unified theory. As a result of the cancellation the one - loop results dominate and we are able to use the one - loop gap equation for the consideration of the chiral symmetry breaking.

 Let us
 apply Fierz transformation to the four fermion term of (\ref{F42}):
\begin{eqnarray}
S_{4}  &=&  \frac{3}{32M_T^2}\int \{ -\theta^2(\bar{\psi}^a\gamma^i
\psi^a)(\bar{\psi}^b\gamma_i \psi^b)\} d^4 x
+ \eta^2(\bar{\psi}^a\gamma^i \gamma^5
\psi^a)(\bar{\psi}^b\gamma_i \gamma^5  \psi^b)\} d^4 x\nonumber\\
&=& \frac{3}{32M_T^2}\int\{4 (\eta^2+\theta^2)
(\bar{\psi}^a_{L}\psi^b_{R})(\bar{\psi}^b_{R} \psi^a_{L})
+(\eta^2-\theta^2)[(\bar{\psi}^a_{L}\gamma_i\psi^b_{L})(\bar{\psi}^b_{L}\gamma^i
\psi^a_{t,L})\nonumber\\&&+(L\leftarrow \rightarrow R)]\} d^4 x \label{S4}
\end{eqnarray}
In this form the action is equal to that of the extended NJL model (see Eq. (4), Eq. (5), Eq. (6) of \cite{ENJL}). In
the absence of the other gauge fields the $SU({ N})_L\otimes
SU({ N})_R$ symmetry is broken down to $SU({N})_V$
(here ${ N}$ is the total number of fermions). We suppose that the Electroweak interactions act
as a perturbation.

 For the purpose of the further consideration we
denote $G_S = \frac{3 (\theta^2+\eta^2)
 \Lambda^2_{\chi}}{64 M_T^2\pi^2}$; $G_V = \frac{\theta^2-\eta^2}{4(\theta^2+\eta^2))}G_S$. Here
$\Lambda_{\chi}$ is the cutoff that is now the physical parameter of the model.
We also denote $g_s =
\frac{4\pi^2G_S}{\Lambda_{\chi}^2}=\frac{3(\theta^2+\eta^2)}{16 M_T^2}$.
Next, the auxiliary fields $H$, $L_i$, and $R_i$ are introduced and the new
action for $\psi$ has the form:
\begin{eqnarray}
S_{4,t}  &=& \int\{ -(\bar{\psi}^a_{t, L}H^+_{ab} \psi^b_{R} + (h.c.)) -
\frac{8 M_T^2}{3 (\theta^2+\eta^2)} \, H_{ab}^+H_{ab}\}d^4x \nonumber\\&&
+\int\{(\bar{\psi}^a_{t,L}\gamma^i L^{ab}_i\psi^b_{t,L})
-\frac{32M_T^2}{3(\theta^2-\eta^2)} {\rm Tr}\,L^i L_{i} +(L\leftarrow \rightarrow R)\}
d^4 x\label{eff}
\end{eqnarray}

Integrating out fermion fields we arrive at the effective action for the
mentioned auxiliary fields (and the source currents for fermion bilinears). The
resulting effective action receives its minimum at $H = m {\bf 1}$, where
$m$ plays the role of the fermion mass (equal for all fermions).
We apply the following regularization: $\frac{1}{-p^2+m^2} \rightarrow \int_{\frac{1}{\Lambda_{\chi}^2}}^{\infty} d\tau
e^{-\tau (-p^2+m^2)}$.
With this regularization the expression for the condensate of $\psi$  is:
\begin{eqnarray}
<\bar{\psi}\psi> &=& i \int\frac{d^4p}{(2\pi)^4}\frac{1}{p\gamma - m}
 =-\frac{1}{16\pi^2}4m^3
\Gamma(-1,\frac{m^2}{\Lambda_{\chi}^2})\label{GM}
\end{eqnarray}

Here $\Gamma(n,x) = \int_x^{\infty}\frac{dz}{z}e^{-z}z^{n}$. The gap equation
is $m = -g_s<\bar{\psi}\psi>$,  that is
\begin{equation}
m = G_S m
\{\exp(-\frac{m^2}{\Lambda_{\chi}^2})-\frac{m^2}{\Lambda_{\chi}^2}
 \Gamma(0,\frac{m^2}{\Lambda_{\chi}^2})\} \approx G_S m (1 - \frac{m^2}{\Lambda^2}{\rm log} \frac{\Lambda_{\chi}^2}{m^2}), \quad (\Lambda_{\chi} \gg m)  \label{gap}
\end{equation}
There exists the critical value of
$G_S$: at $G_S > 1$ the gap equation has the nonzero solution for $m$ while
for $G_S < 1$ it has not. This means that in this approximation the
condensation of fermions occurs at

\begin{equation}
M_T < M_T^{\rm critical} = \sqrt{3 (\theta^2+\eta^2) }\frac{\Lambda_{\chi}}{8\pi}\sim
\Lambda_{\chi}\label{cond}
\end{equation}

The analogue of the technipion decay constant $F_T$ in the given approximation is:
\begin{equation}
F^2_T = \frac{m^2}{8\pi^2}
 \Gamma(0,\frac{m^2}{\Lambda_{\chi}^2}) \approx \frac{m^2}{8\pi^2} \, {\rm log}\, \frac{\Lambda_{\chi}^2}{m^2}, \quad (\Lambda_{\chi} \gg m) \label{FT}
\end{equation}

In order to have appropriate values of $W$ and $Z$ - boson masses we need
$F_T\sim 250/\sqrt{N}$ Gev, where $N=24$ is the total number of fermions.  It is given by
$N = (3 \, {\rm generations}) \times (3\, {\rm  colors} + 1 \, {\rm lepton}) \times (2\, {\rm fermions}\, {\rm  in}\, SU(2) \, {\rm doublets}) = 3\times 4\times 2 = 24$.   Rough estimate via Eq. (\ref{FT}) with $m = 174$ GeV (top quark mass)  gives  the value $\Lambda_{\chi}\sim 5$ TeV. At the same time if only one $t$ quark contributes to the formation of the gauge boson masses we would have $\Lambda_{\chi}\sim 10^{15}$ GeV.
 This, probably, means that Eq. (\ref{FT}) cannot be used directly and should be a subject of a subsequent change.
  In a realistic theory the fermions should have different masses due to the corrections to the gap equations caused by the other interactions (see below). In order to avoid the constraints on the FCNC we need $\Lambda_{\chi} \ge 1000$ TeV.

In the absence of SM interactions the relative orientation of the SM gauge
group $G_W = SU(3)\otimes SU(2)\otimes U(1)$ and $SU({ N})_V$ from
$SU({ N})_L\otimes SU({ N})_R \rightarrow SU({ N})_V$ is
irrelevant. However, when the SM interactions are turned on, the effective
potential due to exchange by SM gauge bosons depends on this relative
orientation. Minimum of the potential is achieved in the true vacuum state and
defines the pattern of the breakdown of $G_W$. This process is known as the vacuum
alignment. In \cite{Align, Align1} this process was considered for the conventional technicolor. It was shown that under very natural suppositions $G_W$
is broken in a minimal way. This means that the subgroups of $G_W$ are not
broken unless they should. We suppose, that the same arguments work also for the dynamical theory of Lorentz group as a source of the dynamical electroweak symmetry breaking. The form of the condensate requires that $SU(2)$ and
$U(1)$ subgroups are broken. That's why in the complete analogy with $SU(N_{TC})$
Farhi - Susskind model the Electroweak group in our case is broken correctly while
the $SU(3)$ group remains unbroken.

\section{Taking into account small perturbations}
\label{NJL}

Let us briefly mention what happens to the gap equation Eq. (\ref{gap}) when the corrections are taken into account due to the Standard Model gauge fields and, due to some other unknown interactions. The four - fermion interaction term of Eq. (\ref{S4}) is modified in this case.  To demonstrate, how this may work, let us consider the toy model that involves only $t$ and $b$ - quarks, and ignore the color. Correspondingly, we have the left - handed doublet $\chi_{A, L}^T = (t_L,b_L)$ and the right - handed doublet $\chi_{A, R}^T = (t_R,b_R)$. The unperturbed action has the form
\begin{eqnarray}
S & = & \int d^4x \Bigl(\bar{\chi}[ i \nabla \gamma ]\chi  +  \frac{8\pi^2}{\Lambda_{\chi}^2} (\bar{\chi}_{A,L}
\chi^{B}_R)(\bar{\chi}_{\bar{B}, R} {\chi}^{{A}}_{L}) I_B^{\bar{B}}\Bigr) \label{toy0}
\end{eqnarray}
with $I_B^{\bar{B}} = \delta_B^{\bar{B}}(1+y)$. The four - fermion terms  similar to the second term of Eq. (\ref{S4}) with the product of two vector - like fermion bilinears do not contribute to the one - loop gap equation. Therefore, we omit these terms. The one - loop gap equation gives equal masses of top and bottom quarks $M = M_t = M_b$:
\begin{equation}
\frac{M^2}{\Lambda^2}{\rm log} \frac{\Lambda^2}{M^2} = y\label{gap00}
\end{equation}

Now let us consider as a perturbation the gauge field $B$ interacting with the right - handed top - quark only.
We imply that the corresponding $U(1)$ symmetry is broken spontaneously, and $B$ receives mass $M_B$ much larger than   $\Lambda_{\chi}$. The corrections to Eq. (\ref{toy0}) due to the exchange by $B$ give the modification of Eq. (\ref{toy0}) with $I_B^{\bar{B}} = {\rm diag} \, \Bigl(1+y_t, 1+y_b\Bigr)$, where $y_b \approx y$, while  $|y_t - y_b| \sim \frac{\Lambda^2_{\chi}}{M_B^2}$.

This toy pattern demonstrates how perturbations may affect action of Eq. (\ref{S4}). Next, we consider the NJL model of general type that generalizes the model with action Eq. (\ref{toy0}) and that involves all $6$ quarks and all $6$ leptons (neutrino is supposed to be of Dirac type).  The particular form of the four  - fermion action is obtained under the supposition that the tensor of coupling constants standing in front of the four - fermion
term is factorized and under the supposition that lepton number originates from the fourth color. At the present moment we do not intend to consider this model as realistic. It is rather an example of how this may work and does not exhaust all possible
deformations of the theory by small perturbations. However, in principle, with certain updates this model may pretend on the description of the TeV - scale physics.  (For the consideration of points, where it is to be updated see the discussion below.) The action of this NJL model is suggested in \cite{VZ2013_2} and generalizes the action suggested in \cite{Miransky} considered in details in \cite{VZ2012,VZ2013}:
\begin{eqnarray}
S & = & \int d^4x \Bigl(\bar{\chi}[ i \nabla \gamma ]\chi  +  \frac{8\pi^2}{\Lambda^2} (\bar{\chi}_{k, \alpha A,L} \chi^{l, \beta, B}_R)(\bar{\chi}_{\bar{l},\bar{{\beta}}
\bar{B}, R} {\chi}^{\bar{k},\bar{\alpha} {A}}_{L}) W^k_{\bar{k}} W_l^{\bar{l}} L_{\bar{\alpha}}^{\alpha}
R_{\beta}^{\bar{\beta}} I_B^{\bar{B}}\Bigr) \label{Stopcolor_}
\end{eqnarray}

Here $\chi_{k, \alpha A}^T = \{(u_k,d_k); (c_k,s_k); (t_k,b_k)\}$ for $k = 1,2,3$ is the set of the quark doublets
with the generation index $\alpha$ while $\chi_{4, \alpha A}^T = \{(\nu_e, e); (\nu_{\mu},\mu); (\nu_{\tau},\tau)\}$ is the set of the lepton doublets.  $\Lambda$ is the dimensional parameter.
Hermitian matrices  $L,R,I,W$  contain dimensionless coupling constants. The form of action Eq. (\ref{Stopcolor_}) with $W = {\rm diag} \, (1+\frac{1}{2}W_{e\mu\tau},1,1,1)$ is fixed by the requirement that there is the $SU(3)\otimes SU(2) \otimes U(1)$ symmetry.
We imply that all eigenvalues of matrices $L,R,I$ are close to each
other.
If all quarks and leptons would experience only the interaction via the gauge bosons of Lorentz group, then in this action the eigenvalues of $L,R,I$ are all equal to each other, and $W_{e\mu\tau}=0$. In this case we would come to the scalar - scalar interaction term of Eq. (\ref{S4}).
Any small corrections to this equality gives the eigenvalues of $L,R,I$ that only
slightly deviate from each other, and the value of $W_{e\mu\tau}$ that only slightly deviates from $0$. (After the suitable rescaling $\Lambda$ plays the role of the cutoff $\Lambda_{\chi}$,
while the eigenvalues of $L,R,I$ are all close to $1$.)
The difference between the model with action of Eq. (\ref{Stopcolor_}) and the model of \cite{VZ2012,VZ2013} is that in the present model the leptons are included and the color indexes (the fourth colour is the lepton number) are contracted in the different way.

The basis of observed quarks corresponds to the diagonal form of $L,R,I$. We
denote
$L = {\rm diag}(1+L_{ude},1+L_{cs\mu},1+L_{tb\tau})$, $R = {\rm
diag}(1+R_{ude},1+R_{cs\mu},1+R_{tb\tau})$,
 $ I = {\rm diag}(1+I_{up},1+I_{down})$, and
\begin{eqnarray}
 &&y_u  = L_{ude} +R_{ude}+ I_{up}, \quad y_d = L_{ude}+ R_{ude}
+I_{down},\nonumber\\&& ...
 \nonumber\\ &&
y_{ud}  = L_{ude} +R_{ude}+ I_{down}, \quad y_{du} = L_{ude} +R_{ude}
+I_{up},\nonumber\\
&& ... \nonumber\\
&&y_{\nu_e}  = L_{ude} +R_{ude}+ I_{up} + W_{e\mu\tau}, \quad y_{e} = L_{ude}+ R_{ude}
+I_{down} + W_{e\mu\tau},\nonumber\\&& ...
 \nonumber\\ &&
y_{\nu_e e}  = L_{ude} +R_{ude}+ I_{down}+W_{e\mu\tau}, \quad y_{e \nu_e} = L_{ude} +R_{ude}
+I_{up}+W_{e\mu\tau},\nonumber\\
&& ...
\label{parameters}
\end{eqnarray}

These coupling constants satisfy the relation $y_{q_1q_2}+y_{q_1q_2} =
y_{q_1}+y_{q_2}$.
As it was mentioned above, it is implied that $|y_{q}|, |y_{q_1 q_2}| << 1$.
Bosonic spectrum of this model is formally given by the expressions for the bosonic spectrum of the model suggested in \cite{Miransky} calculated in one - loop approximation in
\cite{VZ2012,VZ2013}. It is implied that in vacuum the composite scalar
fields $h_q = \bar{q}q$ are condensed for all fermions $q=u,d,c,s,t,b,e,\mu,\tau, \nu_e,\nu_{\mu}, \nu_{\tau}$. The
induced fermion masses $M_q$ are related to the coupling constants $y_q$, $\Lambda$ as
\begin{equation}
\frac{M_q^2}{\Lambda^2}{\rm log} \frac{\Lambda^2}{M_q^2} = y_q\label{gap2}
\end{equation}
(Here it is implied that $\Lambda \gg M_q$.)

Instead of Eq. (\ref{FT}) we have in each of the $q\bar{q}$ channel:
\begin{equation}
F^2_T = \frac{M_q^2}{8\pi^2} \, {\rm log}\, \frac{\Lambda^2}{M_q^2}\label{FT2}
\end{equation}

The $t$ quark dominates. Therefore, in order to have appropriate values of $W$ and $Z$ - boson masses we need
$F_T\sim 250/\sqrt{N}$ Gev, where $N=3$ is the number of colours. This gives $\Lambda = \Lambda_{\chi}\sim 10^{15}$ GeV.

 There are two excitations in each $q\bar{q}$ channel:
\begin{equation}
M^P_{q\bar{q}} = 0; \quad M^S_{q\bar{q}} = 2 M_q
\end{equation}
 and four excitations (i.e. two doubly degenerated excitations) in each
$q_1\bar{q}_2$ channel.  (Pairings of leptons and quarks are also allowed and give the colored scalar fields.) We denote the masses $M^{\pm}_{q_1q_2},
M^{\pm}_{q_2q_1}$. They are given by
\begin{eqnarray}
M_{q_1 {q}_2}^2 &= &M_{q_1}^2 + M_{q_2}^2  \pm
\sqrt{(M_{q_2}^2 - M_{q_1}^2)^2\zeta_{q_1q_2}^2+ 4 M_{q_1}^2M_{q_2}^2}
\end{eqnarray}
with $ \zeta_{q_1 q_2} = \frac{2 y_{q_1q_2}- y_{q_2} - y_{q_1}}{y_{q_2}-y_{q_1}} =
\zeta_{q_2 q_1}$.
It is worth mentioning that each of the scalar quark - antiquark bosons carries two color indexes. In the absence of the $SU(3)$ gauge field each of these channels represents the degenerate nonet. When the color interactions are turned on we are left with the singlet and octet states. Traceless octet states as well as the color scalar excitations of the quark - lepton channels cannot exist as distinct particles due to color confinement.

Instead of the trivial Nambu sum rule of the simplest models of top - quark condensation $M_H = 2 M_t$ we have the sum rule \cite{VZ2013}:
\begin{eqnarray}
&& [M^{+}_{q_1\bar{q}_2}]^2  + [M^{-}_{q_1\bar{q}_2}]^2 +
[M^{+}_{q_2\bar{q}_1}]^2  + [M^{-}_{q_2\bar{q}_1}]^2\approx 4 [M_{q_1}^2 +
M_{q_2}^2],  \quad (q_1\ne q_2); \nonumber\\
&& [M^{P}_{q\bar{q}}]^2  + [M^{S}_{q\bar{q}}]^2 \approx 4 M_{q}^2\label{NSRR}
\end{eqnarray}
In the case when the t - quark contributes to the formation of the given
scalar excitation,
its mass dominates, and in each channel ($t\bar{t}, t\bar{c}$, ...) we come
to the relation $\sum M_{H, i}^2  \approx 4 M^2_t$,
where the sum is over the scalar excitations in the given channel.

We already do not have the only gap equation. Instead, for each fermion particle its own gap equation (\ref{gap2}) appears.
 It is important, that although the corrections to the eigenvalues of $L,R,I,W$ are small, this does not mean that the corrections to the masses are small. Instead, the large difference between masses may appear in this way. We imply that this is provided by the fine - tuning that is responsible for the cancellation of dominant higher - loop divergences in the NJL model. As it was mentioned in the Introduction, this mechanism may be similar to the mechanism that provides the cancellation of the divergent contributions to vacuum energy (due to sound waves) in quantum hydrodynamics.

It is worth mentioning that among the
mentioned Higgs bosons there are 24 Goldstone bosons that are exactly
massless (in the channels $t(1\pm \gamma^5)\bar b, t \gamma^5\bar{t},
c(1\pm\gamma^5)\bar{s}, c\gamma^5\bar{c}$, $u(1\pm \gamma^5)\bar{d},
u\gamma^5\bar{u},b\gamma^5\bar{b}$, $s\gamma^5\bar{s}$, $d\gamma^5\bar{d}$ and $\nu_e(1\pm \gamma^5)\bar e, \nu_e \gamma^5\bar{\nu_e}, \nu_{\mu}(1\pm \gamma^5)\bar \mu, \nu_{\mu} \gamma^5\bar{\nu_{\mu}}, \nu_{\tau}(1\pm \gamma^5)\bar \tau, \nu_{\tau} \gamma^5\bar{\nu_{\tau}}, {\tau} \gamma^5\bar{{\tau}}, {\mu} \gamma^5\bar{{\mu}}, e \gamma^5\bar{e}$).
There are Higgs bosons with the masses of the order of the t-quark mass ($
t(1\pm \gamma^5)\bar b, t \bar{t},  t(1\pm\gamma^5)\bar{s},
t\gamma^5\bar{c}, t(1\pm \gamma^5)\bar{d}, t\gamma^5\bar{u}$ and
$t(1\pm \gamma^5)\bar e, t(1\pm\gamma^5)\bar{\mu},
t(1\pm\gamma^5)\bar{\tau}, t(1\pm \gamma^5)\bar{\nu_e}, t\gamma^5\bar{\nu_{\mu}}, t\gamma^5\bar{\nu_{\tau}}$). The other
Higgs bosons have masses much smaller than the t - quark mass.
If we want to heave realistic model, extra light Higgs bosons should be provided with the masses of
the order of $M_t$. In principle, this may be achieved if the new gauge symmetries are added, that are spontaneously broken. Then the extra light Higgs bosons may become massive via the Higgs mechanism. The consideration of such a completion of the theory is out of the scope of the present paper.

In principle, all neutral Higgs bosons $h^{a}, a = t \bar{t},b \bar{b}, \bar{\tau} \tau, \bar{\nu}_{\tau} \nu_{\tau} ... $ are coupled to the fields of the Standard Model in a similar way. However, already at the tree level the corresponding coupling constants are different for different Higgs bosons.
At the tree level in the effective decay lagrangian \cite{status} the following terms dominate:
\bea
\label{eq:1}
L_{eff}  &=  &
   {2 m_W^2  \over v}  h^{(\bar{t}t)}  \,   W_\mu^+ W_\mu^-  +     {m_Z^2 \over v} h^{(\bar{t}t)}  \,  Z_\mu  Z_\mu
 + c^{}_{g}  {\alpha_s \over 12 \pi v} h^{(\bar{t}t)} \, G_{\mu \nu}^a G_{\mu \nu}^a \nonumber  \\&& +  c^{}_{\gamma} { \alpha \over \pi v} h^{(\bar{t}t)} \, A_{\mu \nu} A_{\mu \nu} -   {m_b \over v^{(\bar{b}b)} } h^{(\bar{b}b)} \,\bar b b  -    {m_c \over v^{(\bar{c}c)}}  h^{(\bar{c}c)} \,   \bar c c \ -   {m_{\tau} \over v^{(\bar{\tau}\tau)} } h^{(\bar{\tau}\tau)} \,\bar \tau \tau .
\eea
We denote by $v^{(a)}$ the vacuum average of the scalar boson in the $a$ - th channel; $v \approx v^{(\bar{t}t)} \approx 245\, {\rm GeV}$ is the vacuum average of the scalar field in the $\bar{t}t$  channel, that dominates the formation of the gauge boson masses.
One can see, that at the tree level in the channels $h \rightarrow gg, \gamma \gamma, ZZ, WW$ the decay of the $\bar{t}t$ Higgs boson dominates. The contributions of the decays of the other Higgs bosons are suppressed. The top quark has been integrated out in Eq.~\eqref{eq:1} and its effects are included in the effective  couplings $c_g$ and $c_\gamma$ that are given by $c_{g}   \simeq 1.03\,, c_{\gamma} =  (2/9)c_{g}  \simeq 0.23$ (see \cite{status}).

One can see, that in the processes like $p p \rightarrow h^{(a)} \rightarrow \bar{c}c, \bar{b}b, \bar{\tau} \tau$ (that may be observed at the LHC) the scalar states $h^{(a)}$ with $a=\bar{c}c$, $\bar{t}{t}$, $\bar{\tau}\tau$ correspondingly dominate at the tree level.
From this consideration it also follows that the cross - sections $\sigma^{(a)}$ of the processes  like $p p \rightarrow h^{(a)} \rightarrow WW, ZZ, gg, \gamma \gamma$ for the $\bar{t} t $ Higgs bosons are much larger than for the other Higgs Bosons and are close to that of the Standard Model.
This means, in particular, that the scalar boson of this model in the $\bar{t}t$ channel with mass $\approx 2 M_t$ is excluded by the LHC data. Therefore, some additional physics is necessary that either suppresses the corresponding cross - section or makes this state much heavier.
The decays of the other neutral  Higgs bosons to $ZZ, WW, \gamma \gamma, gg$ are suppressed compared to that of $\bar{t}t$. Therefore, these scalar states are not excluded by the LHC data.

\section{Lattice discretization}

We suggest the geometrical pattern of manifestly gauge invariant  discretization that is similar to that of the Regge calculus or the pattern suggested in \cite{Z2006} for the discretization of Poincare quantum gravity. It is implied that the Wick rotation is performed. Therefore, we deal with the Euclidean theory and with the gauge group $SO(4)$ instead of $SO(3,1)$. We use the hypercubic form of the lattice cells. We approximate the given manyfold with the $SO(4)$ curvature by the set of the adjacent hypercubes of the same size. It is implied that the Lorentz connection is localized on the sides of these hypercubes. As for the fermion field we consider the approximation, in which inside each hypercube the fermion field is constant. As a result, the lattice sites of the given regularization are placed in the middles of the hypercubes. The $SO(4)$ connection is attached to links. The curvature is localized on the plaquettes. Its values in the adjacent hypercubes are related by the gauge transformation.

The model to be discretized has the action $S = S_f + S_T + S_G$, where the different terms are
given by Eq. (\ref{Sf}), Eq. (\ref{ST}), Eq. (\ref{ST2}).
The lattice discretized action for the fermions is similar to that of suggested in \cite{Diakonov} (we performed the rotation to the Euclidean signature of space - time):
\begin{eqnarray}
S_f & = & \sum_{xy} \{ {\psi}^+_{L,x} H^{xy}_L \psi_{L,y} + \psi^+_{R,x} H^{xy}_R \psi_{R,y} \}, \label{Sfl}
\end{eqnarray}
where
\begin{eqnarray}
H_L & = & \sum_k\Bigl( \eta\{ U^+_{R,yx} \bar{\tau}^k \delta_{x-e_k,y} - \bar{\tau}^k U_{L,xy}  \delta_{x+e_k,y} \}\nonumber\\
&&-i\theta\{ U^+_{R,yx} \bar{\tau}^k \delta_{x-e_k,y} + \bar{\tau}^k  U_{L,xy} \delta_{x+e_k,y} - 2 \bar{\tau}^k \delta_{x,y} \}\Bigr)\nonumber\\
H_R & = & \sum_k \Bigl(\eta\{ U^+_{L,yx} {\tau}^k \delta_{x-e_k,y} - {\tau}^kU_{R,xy}  \delta_{x+e_k,y} \} \nonumber\\
&&-i\theta\{ U^+_{L,yx} {\tau}^k \delta_{x-e_k,y} + {\tau}^k U_{R,xy}  \delta_{x+e_k,y} - 2 {\tau}^k \delta_{x,y} \}\Bigr)\nonumber\\
&& \tau^4 = \bar{\tau}^4 = -i, \quad \tau^a = -\bar{\tau}^a = \sigma^a (a=1,2,3) \label{Sfl_}
\end{eqnarray}
Here the link matrix is
\begin{eqnarray}
U_{yx} &=& \left(\begin{array}{cc} U_{L,yx} & 0 \\
 0 & U_{R,yx} \end{array}\right), \quad U_L,U_R \in SU(2)
 \end{eqnarray}
The action is (formally) invariant under the gauge transformation
\begin{eqnarray}
\psi_x & \rightarrow & G_x \psi_x, \quad \psi_x^+ \rightarrow \psi_x^+\Gamma^4 G_x^+ \Gamma^4, \quad U_{xy} \rightarrow G_x U_{xy} G^+_y, \quad  \tau^k \rightarrow G_L \tau^k G^+_R, \quad \bar{\tau}^k \rightarrow G_R \bar{\tau}^k G^+_L \nonumber\\
G &=& \left(\begin{array}{cc} G_L & 0 \\
 0 & G_R \end{array}\right), \quad G_L,G_R \in SU(2), \quad \Gamma^4 = \left(\begin{array}{cc} 0 & 1 \\
 1 & 0 \end{array}\right)
\end{eqnarray}
This symmetry is broken by the lattice discretization, and is restored in the continuum limit, when the invariance under local $SO(4)$ coordinate  transformations comes back. (The appropriate transformations of the coordinates compensate the transformation of $\tau$ and $\bar{\tau}$.)
The fermion action  can also be rewritten in the following way:
\begin{eqnarray}
S_f & = & \sum_{xy}  {\psi}^+_{x} \Gamma^4 {\cal H}^{xy}_L \psi_{y}, \nonumber\\
{\cal H} &=& \left(\begin{array}{cc} 0 & H_R \\
 H_L & 0 \end{array}\right), \quad H_L^+ = H_R \label{HH}
\end{eqnarray}

Remarkably, for $\theta\ne 0$ in the absence of the gauge field the given discretization gives no doublers just like the Wilson formulation of lattice fermions.

Next, let us introduce the definition of lattice $SO(4)$ curvature. It corresponds to the closed path along the boundary of the given plaquette  started from the given lattice site $x$.  Therefore, it is marked by the position of the lattice site $x$ and the couple of the directions in space - time $n,j$.
\begin{eqnarray}
{\cal G}_{x,n,j}^{4,k}  & = & - {\cal G}_{x,n,j}^{k,4} = i \,{\rm sign}(n)\, {\rm sign}(j) \, \Bigl(  {\rm Tr}\, ( U_{L,x,n,j} - U^+_{L,x,n,j}) \sigma^k  -  {\rm Tr}\, ( U_{R,x,n,j} - U^+_{R,x,n,j}) \sigma^k \Bigr) \nonumber\\
{\cal G}_{x,n,j}^{k,l} & = &   i\, {\rm sign}(n)\, {\rm sign}(j) \, \epsilon^{klm} \Bigl( {\rm Tr}\, ( U_{L,x,n,j} - U^+_{L,x,n,j}) \sigma^m +  {\rm Tr}\, ( U_{R,x,n,j} - U^+_{R,x,n,j}) \sigma^m \Bigr)
\end{eqnarray}
Here  $n,j = \pm 1,\pm 2,\pm 3,\pm 4$. Positive sign corresponds to the positive direction while negative sign corresponds to the negative direction in $4D$ space - time. The plaquette variables obtained by the product of link matrices along the boundary of the plaquette located in $n,j$ plane (starting from the point $x$) are denoted by $U_{R,x,n,j}$ and $U_{R,x,n,j}$. Contraction of indices results in the definition of lattice Ricci tensor $\cal R$ and the lattice scalar curvature $\cal S$. Both these quantities are also attached to the closed paths around the boundaries of the plaquettes and are marked by $x, n, j$.
\begin{eqnarray}
{\cal R}_{x,n,j}^{k}  & = &  {\cal G}_{x,n,j}^{k,|n|}, \quad {\cal S}_{x,n,j}   =    {\cal R}_{x,n,j}^{|j|}
\end{eqnarray}

For the part of the action linear in curvature we have:
\begin{eqnarray}
S_T & = & -\kappa \sum_{x} \sum_{n,j=\pm 1,\pm 2, \pm 3, \pm 4} {\cal S}_{x, n, j}
\end{eqnarray}

Analogue of quantity $A$ from Eq. (\ref{A}) is given by
\begin{eqnarray}
{\cal A}_{x,n,j}  & = &  \sum_{kl}\epsilon^{|n||j|kl} {\cal G}_{x,n,j}^{k,l}
\end{eqnarray}

  Contraction of indices in quadratic expressions results in the following combinations:
\begin{eqnarray}
Q_{x}^{(1)}  & = &  \sum_{k,l, n, j} {\cal G}_{x,n,j}^{|k|,|l|}{\cal G}_{x,n,j}^{|k|,|l|} \nonumber\\
 Q^{(2)}_x &=& \sum_{k,l, n, j} {\cal G}_{x,n,j}^{|k|,|l|}{\cal G}_{x,k,l}^{|n|,|j|}\nonumber\\
 Q^{(3)}_x &=& \sum_{k, n, j} {\cal R}_{x,n,j}^{|k|}{\cal R}_{x,n,j}^{|k|}\nonumber\\
 Q^{(4)}_x &=& \sum_{k, n, j} {\cal R}_{x,n,j}^{|k|}{\cal R}_{x,n,k}^{|j|}\nonumber\\
 Q^{(5)}_x &=& \sum_{n, j} {\cal S}_{x,n,j}{\cal S}_{x,n,j}\nonumber\\
 Q^{(6)}_x &=& \sum_{ n, j} {\cal A}_{x,n,j}{\cal A}_{x,n,j}
\end{eqnarray}
(The sum is over positive and negative values of $n,j,k,l$.)
Finally, the term in the action quadratic in curvature has the form
\begin{eqnarray}
S_G & = & \sum_{x}\sum_{i=1}^6 \beta_i Q^{(i)}_x
\end{eqnarray}


Our main expectation is that in this theory the chiral symmetry is broken dynamically while the confinement is absent.
The main quantities to be measured are the static potential extracted from the Polyakov loops and the chiral condensate.
 The chiral condensate $\chi$ may be calculated using the Banks - Casher relation
\begin{eqnarray}
\chi & = & - \frac{\pi}{V} \langle \nu(0)\rangle,
\end{eqnarray}
where $\nu(\lambda)$ is the density of eigenvalues of the operator $\cal H$ given by Eq. (\ref{HH}), $V$ is the $4D$ volume.

The potential $V_{L\bar{L}}(R)=V_{R\bar{R}}(R)$ between the static (either left - handed or right - handed) fermion and anti - fermion is defined through the relation
\begin{eqnarray}
&&{\rm exp}( - V_{L\bar{L}}(|x-y|) T)  =   \langle {\cal P}_L(x) {\cal P}_R(y) \rangle, \nonumber\\
&& {\cal P}_L(x) = {\rm Tr} \Bigl(\Pi_{K = 0, ... , T-1}U_{L, x + Ke_4, x+(K+1)e_4}\Bigr) , \nonumber\\
&& {\cal P}_R(y) = {\rm Tr} \Bigl(\Pi_{K = 0, ... , T-1}U_{R, x + Ke_4, x+(K+1)e_4}\Bigr)
\end{eqnarray}
Here $x_4 = y_4 = 0$, while $T$ is the lattice extent in time direction.
The potential $V_{L\bar{R}}(R)=V_{R\bar{L}}(R)$ between the static fermion and anti - fermion of different chiralities is given by \begin{equation}
{\rm exp}( - V_{L\bar{R}}(|x-y|) T) = \langle {\cal P}_L(x) {\cal P}_L(y)\rangle = \langle {\cal P}_R(x) {\cal P}_R(y)\rangle
\end{equation}

\section{Conclusions}

In this paper we considered the fermions coupled in a nonminimal way to the gauge theory of Lorentz group. First, we have derived the effective four - fermion low energy theory (the NJL approximation). The derivation was based on the assumption that the couplings $\beta_i$ entering Eq. (\ref{ST2}) are asymptotic free. As a result the term of the gauge field action squared in curvature can be neglected at small enough energies.

In our analysis of the effective NJL low energy theory we follow the proposition of \cite{VZ2012} that the microscopic theory is organized in such a way that the contributions from the trans - $\Lambda_{\chi}$ degrees of freedom cancel the dominant divergences at more than one loop existing in the NJL model. (The unified theory that generalizes the gauge theory of Lorentz group may play the role of such a microscopic theory. We even do not exclude that the gauge theory of Lorentz group itself may play such a role and provide mechanism for this cancellation.) This pattern is similar to that of quantum hydrodynamics \cite{quantum_hydro} and was proposed for the solution of the hierarchy problem and the problem of vacuum energy in \cite{hydro_gravity}.
The one - loop analysis demonstrates that the effective NJL model may provide the chiral symmetry breaking. This results in the dynamical electroweak symmetry breaking and in the appearance of the masses for W,Z bosons, as well as the masses of all existing fermions.

Thus we have suggested the possible new scenario of the dynamical electroweak symmetry breaking. In this scenario Lorentz group plays the role of the technicolor group. But there are no technifermions. The Higgs bosons are composed directly of the existing Standard Model fermions. We have demonstrated that the chiral symmetry breaking may take place in the given model in the approximation when the squared curvature gauge field action $S_G$ is neglected. In the same approximation the confinement is absent. This indicates that in the complete theory with $S_G$ taken into account the chiral symmetry breaking may take place in the absence of confinement.  If this pattern indeed takes place, the chiral condensate provides masses for the electroweak gauge bosons. In the zero order approximation all fermions have equal masses. The fermionic excitations are not confined and may exist as distinct particles. We suggest that the observed  massive Standard Model fermions may appear in this way.

As it was mentioned above, in the zero order approximation all fermion masses are equal to each other. The standard model gauge fields (and, probably, some other unknown fields) interact with the fermions and provide small corrections to the gap equations. In \cite{VZ2013_2} the model was suggested that appears as a result of the deformation of the model, in which the fermions experience only the interaction with the gauge bosons of the Lorentz group. This model is considered in details in section 5 of the present paper. The spectrum of bosonic and fermionic excitations in this model is given by the expressions similar to that of \cite{VZ2012,VZ2013}. We observe, that in this case small corrections to the original model may result in the considerable difference between the fermion masses. This is provided by the mentioned above fine tuning that allows to disregard the higher loops in the NJL approximation. Although the fine tuning of this type is often thought of as unnatural, we do not think it is possible to avoid it in the model that pretends to explain the appearance of the fermion masses from less than $1$ eV for neutrino to above $100$ GeV for the top - quark. The models considered in \cite{VZ2012,VZ2013,VZ2013_2} suffer from various problems that at the present moment do not allow to consider them as realistic. First of all, in the strongly coupled model of this type the spectrum of bosonic excitations  contains scalar excitations with small masses (much smaller than the electroweak scale). One of the possible solutions of this problem was  sketched in \cite{VZ2013}. Namely, these light scalar bosons may appear to be the Goldstone bosons eaten by some gauge fields during the spontaneous breakdown of the additional gauge symmetries. Besides, as it was mentioned in \cite{VZ2013_2}, the scalar state of the $\bar{t}t$ channel is excluded by the present LHC data. So, the mechanism should be added that either suppresses the cross section with the creation of this state, or increases its mass considerably. It is worth mentioning that the model of \cite{VZ2013_2} considered in section 5 of the present paper does not exhaust all possible deformations of the original theory, in which the fermions interact with the gauge field of the Lorentz group. That's why, it could be that more realistic deformations may appear.

We have suggested the lattice discretization of the model, in which fermions interact with the gauge bosons of the Lorentz group. On the lattice the gauge invariance is lost. It is supposed to be restored in the continuum limit. Lattice Dirac operator has several peculiar properties. In the presence of nonminimal interactions the fermion doublers disappear like in the model with Wilson fermions. The Dirac operator is Hermitian, and the Banks - Casher relation can be used in order to check the appearance of the chiral condensate.

\section*{Acknowledgements}
Numerous discussions with G.E.Volovik are kindly acknowledged. The author is greatful to Yu.A.Simonov for the discussion that  stimulated him to consider the unusual scenario for the dynamical electroweak symmetry breaking. The author is especially indebted to D.I.Diakonov for the continuous support during the last two years and for the discussions of issues related to the fermions coupled to gravitational fields. Also the author is greatful to B.Svetitsky for the discussion of the lattice discretization and to V.I.Zakharov for the discussion of the chiral symmetry breaking in the considered model.    This work was partly supported by RFBR grant 11-02-01227, by the
Federal Special-Purpose Programme 'Human Capital' of the Russian Ministry of
Science and Education, by Federal Special-Purpose Programme 07.514.12.4028.


\begin{thebibliography}{99}




\bibitem{CMSHiggs}
"Search for the standard model Higgs boson produced in association with W and Z bosons in pp collisions at "s=7 TeV",  CMS Collaboration
arXiv:1209.3937 ; CMS-HIG-12-010 ; CERN-PH-EP-2012-253.

"Observation of a new boson at a mass of 125 GeV with the CMS experiment at the LHC", CMS Collaboration, arXiv:1207.7235; CMS-HIG-12-028; CERN-PH-EP-2012-220.- Geneva : CERN, 2012 - 59 p. - Published in : Phys. Lett. B 716 (2012) 30-61

"A search for a doubly-charged Higgs boson in pp collisions at "s = 7 TeV",  CMS Collaboration
arXiv:1207.2666 ; CMS-HIG-12-005 ; CERN-PH-EP-2012-169. - 2012. - 39 p.



\bibitem{ATLASHiggs}
"Observation of a new particle in the search for the Standard Model Higgs boson with the ATLAS detector at the LHC",  ATLAS Collaboration,
Phys.Lett. B716 (2012) 1-29, CERN-PH-EP-2012-218,
arXiv:1207.7214 [hep-ex]


\bibitem{CMSexc}
"Observation of a new boson with a mass near 125 GeV", The CMS Collaboration,
CMS PAS HIG-12-020, available at CERN information server as
http://cdsweb.cern.ch/record/1460438/files/HIG-12-020-pas.pdf



\bibitem{ATLASexc}
http://cdsweb.cern.ch/record/1470512/files/ATL-PHYS-SLIDE-2012-459.pdf



\bibitem{VZ2013}
G.~E.~Volovik and M.~A.~Zubkov,
 arXiv:1302.2360, Pisma v ZhETF, vol. 97 (2013), issue 6, page 344

\bibitem{VZ2012}
G.~E.~Volovik and M.~A.~Zubkov,
 Phys. Rev. D 87, 075016 (2013),  arXiv:1209.0204

\bibitem{VZ2013_2} G.E.Vlovik, M.A.Zubkov, 	arXiv:1305.7219, to appear in Journal of Low temperature Physics



\bibitem{Technicolor}
 Christopher T. Hill, Elizabeth H. Simmons,
 Phys.Rept. 381 (2003) 235-402;
Erratum-ibid. 390 (2004) 553-554

\bibitem{Technicolor_}
Kenneth Lane, hep-ph/0202255

\bibitem{Technicolor__}
R. Sekhar Chivukula, hep-ph/0011264

\bibitem{Simonov}
 Yu.A.Simonov, "On dynamics of fermion generations", arXiv:0912.1946, Phys.Atom.Nucl.73:1893-1907,2010

Yu.A.Simonov, "Coherent mixing in three and four quark generations",  arXiv:1004.2672, Phys.Atom.Nucl.74:643-649,2011

Yu. A. Simonov, "Spontaneous SU(2) symmetry violation in the $SU(2)_L \times SU(2)_R\times SU(4)$ electroweak model",
arXiv:1012.1171

\bibitem{cveticreview}
G.Cvetiˇc, Rev. Mod. Phys. 71, 513 (1999), arXiv hep-ph/9702381

\bibitem{Minkowski}
Nakia Carlevaro, Orchidea Maria Lecian, Giovanni Montani, Int. J. Mod. Phys. A
23, 1282-1285 (2008)

\bibitem{Minkowski_}
Nakia Carlevaro, Orchidea Maria Lecian, Giovanni Montani,
Mod.Phys.Lett.A24:415-427,2009





\bibitem{topcolor1}
William A. Bardeen, Christopher T. Hill, Manfred Lindner,
"Minimal Dynamical Symmetry Breaking of the Standard Model,"
Phys. Rev. D {\bf 41}, 1647--1660 (1990).

\bibitem{topcolor}
Elizabeth H. Simmons, R. Sekhar Chivukula, Baradhwaj Coleppa, Heather E. Logan, Adam Martin,
"Topcolor in the LHC Era",
  arXiv:1112.3538

%
\bibitem{Marciano:1989xd}
  W.~J.~Marciano,
  Phys.\ Rev.\ Lett.\  {\bf 62}, 2793 (1989).

\bibitem{Hill:1996te}
  C.~T.~Hill,
  arXiv:hep-ph/9702320.

\bibitem{Buchalla:1995dp}
  G.~Buchalla, G.~Burdman, C.~T.~Hill, D.~Kominis,
  Phys.\ Rev.\  {\bf D53}, 5185-5200 (1996).
  [hep-ph/9510376].


\bibitem{topcolor}
Elizabeth H. Simmons, R. Sekhar Chivukula, Baradhwaj Coleppa, Heather E. Logan, Adam Martin,
"Topcolor in the LHC Era",
  arXiv:1112.3538

\bibitem{modern_topcolor}
R. Sekhar Chivukula, Baradhwaj Coleppa, Pawin Ittisamai, Heather E. Logan, Adam Martin, Jing Ren, Elizabeth H. Simmons, "Discovering Strong Top Dynamics at the LHC", arXiv:1207.0450



\bibitem{Weinberg:1979bn}
  S.~Weinberg,
  Phys.\ Rev.\  D {\bf 19}, 1277 (1979).

\bibitem{Susskind:1978ms}
  L.~Susskind,
  Phys.\ Rev.\  {\bf D20}, 2619-2625 (1979).


\bibitem{Chivukula:1990bc}
  R.~S.~Chivukula, A.~G.~Cohen and K.~D.~Lane,
  Nucl.\ Phys.\  B {\bf 343}, 554 (1990).



\bibitem{Hill:1994hp}
  C.~T.~Hill,
  Phys.\ Lett.\  B {\bf 345}, 483 (1995)
  [arXiv:hep-ph/9411426].

\bibitem{LaneEichten}
K. Lane and E. Eichten,
Phys. Lett. \textbf{B 352}: 382-387 (1995).

\bibitem{Popovic:1998vb}
  M.~B.~Popovic and E.~H.~Simmons,
  Phys.\ Rev.\  D {\bf 58}, 095007 (1998)
  [arXiv:hep-ph/9806287].

\bibitem{Braam:2007pm}
  F.~Braam, M.~Flossdorf, R.~S.~Chivukula, S.~Di Chiara and E.~H.~Simmons,
  Phys.\ Rev.\ {\bf 77}, 055005 (2008)
  [arXiv:0711.1127 [hep-ph]].





\bibitem{Chivukula:2011ag}
  R.~S.~Chivukula, E.~H.~Simmons, B.~Coleppa, H.~E.~Logan, A.~Martin,
  Phys.\ Rev.\  {\bf D83}, 055013 (2011).
  [arXiv:1101.6023 [hep-ph]].




\bibitem{top_coloron}
 R. Sekhar Chivukula, Elizabeth H. Simmons, Natascia Vignaroli, "A Flavorful Top-Coloron Model", arXiv:1302.1069

\bibitem{top_coloron1}
Joshua Sayre, Duane A. Dicus, Chung Kao, S. Nandi, Phys.Rev.D84:015011,2011

\bibitem{topseesaw}
R. Sekhar Chivukula, Bogdan A. Dobrescu, Howard Georgi, Christopher T. Hill,
Phys.Rev.D59:075003,1999


\bibitem{Miransky}
V.A. Miransky,  Masaharu Tanabashi, Koichi Yamawaki,
``Dynamical electroweak symmetry breaking with large anomalous
dimension and t quark condensate",
Phys. Lett. B {\bf 221}, 177--183 (1989).

``Is the t quark responsible for the mass of W and Z bosons?",
Mod.  Phys. Lett. A {\bf 4}, 1043--1053  (1989).

\bibitem{NJL}
Y. Nambu, G. Jona-Lasinio,
"Dynamical model of elementary particles based on an analogy with superconductivity. I,"
Phys. Rev. {\bf 122}, 345--358 (1961).





\bibitem{topcolor1}
William A. Bardeen, Christopher T. Hill, Manfred Lindner,
"Minimal Dynamical Symmetry Breaking of the Standard Model,"
Phys. Rev. D {\bf 41}, 1647--1660 (1990).



\bibitem{cvetic}
G. Cvetic, E. A. Paschos, and N. D. Vlachos, Phys. Rev. D 53 (1996), 2820;
G.Cvetic, Annals of physics 255, 165-203 (1997)

\bibitem{NJLQCD}
M.K. Volkov and A. Radzhabov, Phys. Usp. 49, 551 (2006).


\bibitem{Simmons}
Christopher T. Hill, Elizabeth H. Simmons,
"Strong Dynamics and Electroweak Symmetry Breaking",
Phys. Rept. {\bf 381}, 235--402 (2003) ; Erratum-ibid. {\bf 390}, 553--554 (2004).




\bibitem{latticeNJL}
Costas Strouthos, Stavros Christofi, "Monte Carlo simulations of the NJL model near the nonzero temperature phase transition",
 JHEP0501:057,2005

\bibitem{quantum_hydro}
G.E. Volovik, "From Quantum Hydrodynamics to Quantum Gravity", arXiv:gr-qc/0612134,
 Proceedings of the Eleventh Marcel Grossmann Meeting on General Relativity, edited by H. Kleinert, R.T. Jantzen and R. Ruffini, World Scientific, Singapore, 2008, pp. 1404-1423

\bibitem{hydro_gravity}
G.E. Volovik, "Vacuum energy: quantum hydrodynamics vs quantum gravity", arXiv:gr-qc/0505104,  JETP Lett. 82 (2005) 319-324; Pisma Zh.Eksp.Teor.Fiz. 82 (2005) 358-363



\bibitem{NJLconf}
Robert S. Plant, Michael C. Birse,  Nucl.Phys. A628 (1998) 607-644


\bibitem{Freidel:2005sn}
L.~Freidel, D.~Minic and T.~Takeuchi,  Phys.\ Rev.\  D {\bf 72} (2005) 104002
[arXiv:hep-th/0507253].




\bibitem{Randono:2005up}
A.~Randono, arXiv:hep-th/0510001.


\bibitem{Mercuri:2006um}
S.~Mercuri,  Phys.\ Rev.\  D {\bf 73} (2006) 084016 [arXiv:gr-qc/0601013].

\bibitem{Alexandrov}
 Sergei Alexandrov,
Class.Quant.Grav.25:145012,2008

\bibitem{Xue}
She-Sheng Xue, Phys.Lett.B665:54-57,2008, ArXiv:0804.4619

\bibitem{Alexander1}
S.Alexander, T.Biswas, G.Calcagni, Phys. Rev. D 81, 043511 (2010),
ArXiv:0906.5161

\bibitem{Alexander2}
S.Alexander, D.Vaid, ArXiv:hep-th/0609066


\bibitem{Z2010_3}
M.A.Zubkov, Mod. Phys. Lett. A25:2885-2898,2010, ArXiv:1003.5473





\bibitem{Shapiro}
A.S.Belyaev, I.L.Shapiro, Nucl.Phys. B543 (1999) 20-46, ArXiv:hep-ph/9806313





\bibitem{ExtendedTechnicolor}
 Thomas Appelquist, Neil Christensen, Maurizio Piai, Robert Shrock, Phys.Rev. D70 (2004) 093010

\bibitem{ExtendedTechnicolor_}
 Adam Martin, Kenneth Lane, Phys.Rev. D71 (2005) 015011

\bibitem{ExtendedTechnicolor__}
 Thomas Appelquist, Maurizio Piai, Robert Shrock, Phys.Rev. D69
(2004) 015002

\bibitem{ExtendedTechnicolor___}
Robert Shrock, hep-ph/0703050

\bibitem{ExtendedTechnicolor____}
 Adam Martin, Kenneth Lane, Phys.Rev. D71 (2005) 015011

\bibitem{walking}

Thomas Appelquist, Anuradha Ratnaweera, John Terning, L. C. R. Wijewardhana,
Phys.Rev. D58 (1998) 105017


\bibitem{minimal_walking}
R. Foadi, M.T. Frandsen, T. A. Ryttov, F. Sannino, arXiv:0706.1696

\bibitem{minimal_walking_}
Sven Bjarke Gudnason, Chris Kouvaris, Francesco Sannino,   Phys.Rev. D73 (2006)
115003

\bibitem{minimal_walking__}
D.D. Dietrich (NBI), F. Sannino (NBI), K. Tuominen, Phys.Rev. D72 (2005) 055001


\bibitem{Kagan}
Alex Kagan,  CCNY-HEP-91-12,  Proc. of 15th Johns Hopkins Workshop on Current
Problems in Particle Theory, Baltimore, MD, Aug 26-28, 1991,
 Johns Hopkins Wrkshp 1991:217-242
(QCD161:J55:1991)

\bibitem{Dobrescu_Kagan}
Bogdan A. Dobrescu, Nucl.Phys.B449:462-482,1995

D.Atwood, A.Kagan and T.G.Rizzo,
Phys.\ Rev.\ D {\bf 52}, 6264 (1995) [arXiv:hep-ph/9407408].

A.L.Kagan,
Phys.\ Rev.\ D {\bf 51}, 6196 (1995) [arXiv:hep-ph/9409215].


B.A.~Dobrescu and E.H.Simmons,
Phys.\ Rev.\ D {\bf 59}, 015014 (1999) [arXiv:hep-ph/9807469].


B.A.Dobrescu and J.Terning,
Phys.\ Lett.\ B {\bf 416}, 129 (1998) [arXiv:hep-ph/9709297].


\bibitem{Diakonov}
A.~A.~Vladimirov and D.~Diakonov,
{ ``Phase transitions in spinor quantum gravity on a lattice''},
  {}arXiv:1208.1254 [hep-th]
  {}10.1103/PhysRevD.86.104019,
{}Phys.\ Rev.\ D {\bf 86}, 104019 (2012) 

{}D.~Diakonov,
{ ``Towards lattice-regularized Quantum Gravity''},
  {}arXiv:1109.0091 [hep-th]
\bibitem{Z2006}
M.A.Zubkov, { ``Gauge invariant discretization of Poincare quantum gravity''},
 {}Phys.\ Lett.\  B {\bf 638}, 503 (2006),
  [Erratum-ibid.\  B {\bf 655}, 307 (2007)]
  [arXiv:hep-lat/0604011]



\bibitem{ENJL}
J.Bijnens, C.Bruno, E. de Rafael, Nucl.Phys. B390 (1993) 501-541, hep-ph/920623

\bibitem{Diakonov2}
D.~Diakonov, A.~G.~Tumanov and A.~A.~Vladimirov,
{ ``Low-energy General Relativity with torsion: A Systematic derivative expansion''},
{}arXiv:1104.2432 [hep-th], {}10.1103/PhysRevD.84.124042, {}Phys.\ Rev.\ D {\bf 84}, 124042 (2011) 


\bibitem{ConformalGrav}
 Philip D. Mannheim, Prog.Part.Nucl.Phys. 56 (2006) 340-445

V.V. Zhytnikov, Int.J.Mod.Phys.A8:5141-5152,1993.


\bibitem{Sesgin}
E. Sezgin, P. van Nieuwenhuizen, Phys.Rev.D22:301,1980.

\bibitem{Elizalde}
E. Elizalde, S.D.Odintsov, Int.J.Mod.Phys.D2:51-58,1993

\bibitem{Align}
J. Preskill, Nucl. Phys. B177, 21 (1981)

\bibitem{Align1}
 M. E. Peskin, Nucl. Phys. B175, 197
(1980).




\bibitem{status}"Higgs After the Discovery: A Status Report",
Dean Carmi, Adam Falkowski, Eric Kuflik, Tomer Volansky, Jure Zupan, arXiv:1207.1718



\end{thebibliography}
\end{document}